# Constraining the Dust Coma Properties of Comet C/Siding Spring (2013 A1) at Large Heliocentric Distances

Jian-Yang Li (李荐扬)[1], Nalin H. Samarasinha[1], Michael S. P. Kelley[2], Tony L. Farnham[2], Michael F. A'Hearn[2], Max J. Mutchler[3], Carey M. Lisse[4], W. Alan Delamere[5]

[1] Planetary Science Institute, 1700 E. Ft. Lowell Rd., Suite 106, Tucson, AZ 85719, USA; jyli@psi.edu, nalin@psi.edu
[2] Department of Astronomy, University of Maryland, College Park, MD 20742, USA; msk@astro.umd.edu, farnham@astro.umd.edu, ma@astro.umd.edu
[3] Space Telescope Science Institute, 3700 San Martin Drive, Baltimore, MD 21218-2463, USA; mutchler@stsci.edu
[4] Johns Hopkins University Applied Physics Laboratory, Space Department, 11100 Johns Hopkins Rd., Laurel, MD 20723, USA; carey.lisse@jpuapl.edu
[5] Delamere Support Service, Boulder, CO 80304, USA; alan@delamere.biz

**Abstract:**
The close encounter of Comet C/2013 A1 (Siding Spring) with Mars on October 19, 2014 presented an extremely rare opportunity to obtain the first flyby quality data of the nucleus and inner coma of a dynamically new comet. However, the comet's dust tail potentially posed an impact hazard to those spacecraft. To characterize the comet at large heliocentric distances, study its long-term evolution, and provide critical inputs to hazard modeling, we imaged C/Siding Spring with the *Hubble Space Telescope* when the comet was at 4.58, 3.77, and 3.28 AU from the Sun. The dust production rate, parameterized by the quantity $Af\rho$, was 2500, 2100, and 1700 cm (5000-km radius aperture) for the three epochs, respectively. The color of the dust coma is 5.0±0.3%/100 nm for the first two epochs, and 9.0±0.3%/100 nm for the last epoch, and reddens with increasing cometocentric distance out to ~3000 km from the nucleus. The spatial distribution and the temporal evolution of the dust color are most consistent with the existence of icy grains in the coma. Two jet-like dust features appear in the north-northwest and southeast directions projected in the sky plane. Within each epoch of 1-2 hour duration, no temporal variations were observed for either feature, but the PA of the southeastern feature varied between the three epochs by ~30°. The dust feature morphology suggests two possible orientations for the rotational pole of the nucleus, (RA, Dec) = (295°±5°, +43°±2°) and (190°±10°, 50°±5°), or their diametrically opposite orientations.

## 1. Introduction

Comet C/2013 A1 (Siding Spring) is unique. It is a dynamically new (DN) comet (MPEC 2014-S34) that passed by Mars at a distance of 140,246 km on October 19, 2014 (JPL orbit solution 101). This encounter distance is about 1/16$^{th}$ that of the closest known comet approach to the Earth (Sekanina & Yeomans 1984). The HiRISE camera (McEwen et al. 2007) aboard the Mars Reconnaissance Orbiter resolved the nucleus at down to 140 m/pixel resolution.

High-resolution images from cometary flyby missions have fundamentally revolutionized our understanding about comets (e.g., A'Hearn 2011). However, all the target comets have been periodic comets, and likely have undergone considerable evolution in their physical properties at least for the surfaces and top tens of meters due to repeated solar heating near the Sun (e.g. Belton 2010; Cheng et al. 2013). DN comets, on the other hand, are entering the inner solar system for the first time since being scattered into the Oort cloud after their formation, and stayed there for the age of the Solar System (e.g. Dones et al. 2004). Therefore, DN comets are expected to have much more primitive properties than periodic comets, although their surfaces might have been altered by exposures to cosmic particles (Stern 1990). In addition, new dynamical (Tsiganis et al. 2005; Gomes et al. 2005; Morbidelli et al. 2005; Walsh et al. 2011) and observational studies (A'Hearn et al. 2012; Hartogh et al. 2011) are also challenging the classic view of the formation regions of periodic and Oort cloud comets (e.g., Levison 1996). However, the unpredictability and single passage in the inner solar system for DN comets make flyby missions extremely challenging. Therefore, the close encounter of comet C/Siding Spring with Mars is an extraordinary opportunity to observe a DN comet from a close distance for the first time. It is also a rare opportunity to study the interactions between a cometary coma and the Martian atmosphere (e.g. Yelle et al. 2014).

On the other hand, the dust grains in the coma and along the orbit of C/Siding Spring could potentially pose impact hazards to Mars orbiting spacecraft due to the high encounter velocity of 56 km s$^{-1}$. Numerical simulations suggested that the impact hazard is sensitive to the amount of mm or larger dust grains in the coma, their ejection velocity, and the time when they are ejected (Tricarico et al. 2014; Kelley et al. 2014; Farnocchia et al. 2014; Ye & Hui 2014). The dust activity of the comet one to several years before the Mars encounter is the key to determine the impact hazards.

## 2. Observations and Data Reduction

We observed C/Siding Spring with the *Hubble Space Telescope* (*HST*) Wide Field Camera 3 (WFC3) in three epochs (Table 1). The first epoch used two consecutive *HST* orbits covering two hours in duration; the other two epochs used one orbit each. The January 2014 epoch was specifically chosen when the Earth (and *HST*) crossed the orbital plane of the comet, a viewing geometry advantageous for searching for large grains, and measuring the ejection velocity of dust grains. We adopted similar observing strategy and filters as the observations of comet C/2012 S1 (ISON) in Li et al. (2013). The pixel scale of WFC3/UVIS is 0.04". The images of the comet are shown in Fig. 1. Readers are referred to Li et al. for the details of the data reduction.

## 3. Analysis and Results
*3.1 Photometry*

Table 1 lists the total flux, magnitude, and the $Af\varrho$ (the product of albedo, dust filling factor, and aperture size is a proxy for dust production rate; A'Hearn et al., 1984) of C/Siding Spring. The sky background is estimated from three regions in the sunward direction and two sides of the tail, all outside of the coma and tail. We observed no temporal brightness variations >0.03 mag peak-to-peak photometric scatter among 8-10 images in each epoch, allowing us to average the measurements for each filter in each epoch. The statistical uncertainty in photometry is <<1% for F606W images, and ≤1% for F438W images, and the absolute photometric uncertainty of ~5% dominates the total uncertainty. We corrected $A(\theta)f\varrho$ to zero-phase angle values using the phase function of dust (Marcus 2007; Schleicher & Bair 2011; http://asteroid.lowell.edu/comet/dustphase.html), with the correction factors 0.66, 0.59, and 0.61, respectively.

The azimuthally averaged surface brightness of the coma is close to, but slightly steeper than, $1/\varrho$ (Table 1), where $\varrho$ is cometocentric distance, for $\varrho$<5000 km, suggesting that the dust in the coma is close to a constant speed expansion model. The slightly steeper coma brightness falloff in the F438W filter than the F606W filter suggests reddening of the coma with cometocentric distance (see §3.2).

*3.2 Dust coma color*

We calculated the color slope in concentric annuli centered at the nucleus from the F606W and F438W filters (Fig. 2a). For the inner coma with $\varrho$<1000 km, the color slope is ~3%/100 nm, and did not change over the three epochs. For 1000<$\varrho$<3000 km, the slope increases with distance. Beyond ~3000 km, the color stayed constant until ~14,000 km, outside of which the signal in the F438W images is too low for a reliable color measurement. Comparing the three epochs, the color of the dust coma for 3000<$\varrho$<14,000 km increased from 5.0±0.3%/100 nm to 9.0±0.3%/100 nm. The annulus color trend of C/Siding Spring is consistent with the maps of its coma color slopes (Fig. 2b). Comparing with the morphology of the jet-like features in the inner coma (see §3.4), some weak correlation between the relatively bluer region and the features is evident.

Compared to the color slope of 5.0%±0.2%/100 nm for comet C/ISON measured with a similar-sized aperture at 4.15 AU heliocentric distance, the dust color of C/Siding Spring is similar at 4.58 AU, but slightly redder at 3.77 and 3.28 AU. Compared to most periodic comets with typically 10-15%/100 nm color slopes measured from ground-based observations, C/Siding Spring is relatively bluer, and consistent with a few DN comets that have been observed (Jewitt & Meech 1986; Lowry et al. 1999; Hadamcik & Levasseur-Regourd 2009). The heliocentric distance dependence of cometary dust color was not detected previously (Jewitt & Meech 1988; Solontoi et al. 2012), although those studies were dominated by periodic comets, and used photometric apertures with the same angular size (thus varying physical sizes at comets for different observer distances).

The spatial distribution of the color of C/Siding Spring's dust coma is similar to that of C/ISON at a similar spatial scale (Li et al. 2013) and comet C/1995 O1 (Hale-Bopp) at a much larger spatial scale (Weiler et al. 2003). Li et al. interpreted the spatial color distribution of C/ISON as an indication of the existence of water ice grains in the coma. As the icy grains move away from the nucleus, they slowly sublimate and cause reddening of the coma with cometocentric distance. Similar interpretation applies to C/Siding Spring observed here. Near-infrared spectra from NASA's *Infrared Telescope Facility* also suggest the existence of icy grains in the coma of C/Siding Spring at similar heliocentric distances. We do note that, as with the similar observations of C/ISON, emission from any $C_3$ in the coma could potentially contaminate the F438W filter images (cf. Feldman et al. 2004), although at >3 AU the contamination should be less than a few percent.

If icy grains are causing the blueness, then the possible correlation of blueness with jet-like features indicates that those features contain more abundant icy grains than the ambient coma. The Deep Impact flyby (DIF) spacecraft imaged the icy grains in the inner coma of comet 103P/Hartley 2 (103P hereafter) during its flyby in 2010 (A'Hearn et al. 2011; Kelley et al. 2013), and showed a strong correlation between $CO_2$ jets and a concentration of icy grains (Feaga et al. 2012; Protopapa et al. 2014). The DIF observations also revealed that $CO_2$ is the main driver for the dust activity in the vicinity of the nucleus of 103P even at 1 AU heliocentric distance. During our observations, C/Siding Spring was at >3.2 AU heliocentric distance, where CO and/or $CO_2$ likely drove the bulk of its activity. The observations by *NEOWISE* in January 2014 and *Spitzer Space Telescope* in March 2014 confirmed the existence of a likely $CO_2$-dominated gas coma around C/Siding Spring (Farnham et al. 2014). Therefore, our observations are consistent with the 103P scenario, with icy grains ejected from the nucleus by $CO/CO_2$ outgassing. The icy grains sublimate while they retreat from the nucleus, producing the observed spatial and temporal color variations. Note that the activity level of 103P and its heliocentric distance during the DIF flyby were substantially different from the case of C/Siding Spring, precluding a quick quantitative comparison. The examples of C/Siding Spring and C/ISON suggest that ejection of icy grains by $CO/CO_2$ outgassing might be a common mechanism for water ice sublimation in DN comets while they are outside of 3 AU from the Sun.

*3.3 The size and velocity of dust grains*

The tail of C/Siding Spring showed obvious curvature at October 2013 and March 2014 epochs, due to the effects of $\mu$m-sized grains under the influence of radiation pressure (Burns et al. 1979). We calculated the zero initial velocity syndynes (Finson & Probestein 1968) for C/Siding Spring at all three observing epochs, and overlaid them on the corresponding images (Fig. 3). Syndynes mark the loci of dust grains with the same size but different ejection times. The Earth crossed the orbital plane of C/Siding Spring on January 21, 2014, aligning all the syndynes with the orbital plane as projected on the sky. Inspection of the January C/Siding Spring images shows that the tail deviates to the north of the orbital plane. Since radiation pressure has no influence perpendicular to the plane, such deviation suggests that dust emission is not isotropic, but preferentially towards north. Taking into account this effect, using the syndynes as a general guide in the images from the other two epochs, we conclude that the tail morphology of C/Siding Spring is dominated by dust grains with $\beta$ (the ratio between radiation pressure and solar gravity) of 0.1-0.01, corresponding to grains ~1-10 $\mu$m in size (Burns et al. 1979). The preferential direction in the comet's dust emission could have some effect on the details of this simple interpretation (e.g. Lisse et al. 1998, 2004), though the general results are supported by more detailed modeling (Farnham et al. 2014).

The dust grains ejected in the sunward direction travel outward until they are turned back by solar radiation pressure. The standoff distance of the sunward edge of the coma is determined by the grain size and ejection velocity. Using the simple model outlined by Farnham & Schleicher (2005) and Mueller et al. (2013), the quantity $v^2/\beta$ can be constrained by the standoff distance, where $v$ is the ejection velocity of dust grains. The ejection velocities of dust grains in C/Siding Spring's coma are calculated to be 3.6, 5.9, 6.9 m/s for the three epochs, respectively, assuming $\beta$=0.01. The uncertainty is dominated by the measurement of the coma standoff distance, and should be about 10%. The grain size and velocity derived from this simplified models are supported by the more detailed modified Finson-Probstein modeling with the full dust tail morphology of C/Siding Spring (Farnham et al. 2014).

*3.4 Coma morphology and rotational pole*

After the $1/\varrho$ enhancement, two jet-like features in C/Siding Spring's coma are clearly visible at all three epochs, one extending towards the northwest in the sky plane, and the other towards the south-southeast (Fig. 1). The projected directions of the two features remained nearly unchanged over the three epochs, although the projected solar directions and orbital velocity vectors moved dramatically, suggesting that cometary activity rather than solar radiation pressure or the comet's orbital motion dominate the projected directions of these features. The northwestern feature appears relatively narrow with well-defined edges, while the southeastern feature is broader with more diffusing edges. Interestingly, the northwestern feature appeared to be much brighter than the southeastern feature in October 2013, of comparable brightness in January 2014, and much fainter in March 2014.

With no variations in either the brightness or the morphology of either feature observed within each epoch, it is possible that both features originate from high-latitude regions on the nucleus, and the apparent PAs of the features denote the projected rotational axis of the nucleus. If the PAs do define the pole direction and the pole is fixed in time, then the changing geometry should reveal the position of the axis in inertial space (e.g. Farnham & Cochran, 2002). The PAs of both features in all three epochs are measured by fitting the azimuthal brightness profile 5-20 pixels from the photocenter with a 3$^{rd}$ order polynomial (Table 1). Fig. 4a shows the projections of the half-great circle solutions that would produce the observed PAs. The half great circles corresponding to the northwestern feature intersect within a narrow strip with RA=295°±5°, Dec=+43°±2°, defining the rotational axis (or its diametrical opposite) of the nucleus (solution #1 hereafter). The obliquity of this pole is 51°, and the sub-Earth latitudes for this pole solution are 60°, 38°, and 29° in the three epochs, respectively. Thus in October 2013, the northwestern feature would have been more highly projected, pointing closer to the Earth than in January and March 2014. Such a projection effect might make the feature appear brighter in October than later. On the other hand, however, the half great circles for the southeastern feature do not intersect in one common region in the sky, and cannot easily explain the southeastern feature. The inconsistency may indicate that the southeastern feature originates from an extended area at low latitude, and its brightness and orientation are modulated by the rotation of the nucleus that are too slow to have observational effects within any single epoch. Another possibility is that the southeastern feature is the combined effect of many small features originated from multiple sources, similar to the case of 81P/Wild 2 and 9P/Tempel 1 (Farnham & Schleicher 2005; Sekanina et al. 2004; Farnham et al. 2007; 2013).

Alternatively, since the two coma features are almost on opposite sides of the nucleus, we could also assume that both features are associated with a single source region on the nucleus at low latitude, with the rotation pole perpendicular to the feature PAs. In this scenario, the dust outflow from such a source sweeps around both sides of the nucleus as the nucleus rotates, and the pole would need to lie within ~10° to the sky plane for the two features to be well defined. As before, the pole can be derived using the corresponding great circles solutions at all three epochs. Fig. 4b shows that the pole is constrained in a strip within RA=190°±10°, Dec=+50°±5° (solution #2 hereafter), with an obliquity of 83°. However, the sub-Earth latitudes based on this pole solution are 38°, 62°, and 60° for the three epochs, respectively, which is inconsistent with the assumption that the axis should lie near the sky plane. In addition, the sub-Earth latitude only changed by ~2° from January to March 2014, while the morphologies of the coma features in those two epochs exhibited dramatic changes.

**4. Discussion**

Our *HST* observations do not reveal any temporal variations in the coma potentially associated with the rotation of the nucleus. Several possible explanations exist. First, since the longest duration of our observing epochs is ~2 hours, these observations are insensitive to rotational periods >8 hours. Second, if the comet is emitting dust near its rotational pole as assumed for pole solution #1, or multiple sources distributed along the longitude, then rotational modulation in the brightness of comet is not expected. Third, the combination of low dust emission velocity, velocity dispersion, relatively fast nuclear rotation, and large geocentric distance could smear out the temporal variation in the coma (Samarasinha et al. 1997).

The spatial change of the color with respect to cometocentric distance and the temporal reddening as defined by the *HST* observations contain clues about the abundance of ice in the coma. The lifetimes of solid, pure water ice grains 0.1-100 $\mu$m in size at 3-4 AU heliocentric distance are well over $10^9$ s (Lien 1990). For the derived coma outflow speeds, pure ice grains should fully sublimate on >$10^6$ km length-scales, orders of magnitude larger than our observed icy coma. Solid ice grains mixed with a small amount of dust have lifetimes of the order $10^3$ s and larger at 3.7 AU (Beer et al. 2006), corresponding to >10 km lengthscales. Therefore, the water ice comae of C/Siding Spring and C/ISON are more compatible with dirty water ice. More thorough simulations designed to reproduce the observations are needed to confirm the icy grain interpretations of color, and to understand the behavior of coma ice at large heliocentric distances.

The two possible pole solutions result in very different rotation geometries of the nucleus during its Mars encounter. If the pole is near solution #1, then the sub-solar and sub-Mars latitudes are at ~25° on the opposite sides of the equator pre-encounter, with the sub-Mars point moving to the same side at similar latitude as the sub-solar point post-encounter. If the pole is near solution #2, then the geometry is the opposite, where sub-solar and sub-Mars points are on the same side of the equator pre-encounter, moving to the opposite sides post-encounter. The sub-solar latitude for both cases are similar, but the sub-Mars latitude is ~±50° for solution #2, much higher than that for solution #1 both pre- and post-encounter. The obvious difference in the geometry means that, if there are dust features visible in the inner coma showing any morphological changes caused by nucleus rotation in the time frame of a day, then the observations from Mars could distinguish the two cases and fully determine the rotational pole.

The two pole solutions also result in significantly different seasonal effects on the nucleus. For solution #1, the sub-solar latitude decreases from +50° during pre-perihelion to -30° near perihelion, reaching -50° in December 2014 before moving back to lower latitude. The most rapid change in sub-solar altitude is from August to December 2014. We can expect new active areas developing as the sub-solar point moves. For solution #2, the sub-solar latitude increases from +25° pre-perihelion to +80° in late July 2014, and then rapidly moves back to +30° near perihelion, and continues to cross the equator towards -80° post-perihelion. For this solution, the Sun will not illuminate new area until just past perihelion. The different seasonal effects between the two pole solutions might result in different observational effects on the brightness of the comet, and/or its coma morphology near perihelion.

The *HST* observations of C/Siding Spring place strong constraints on the ejection velocity of dust, its spatial distribution, and the production rate. Based on the *HST* observations and many other observational datasets, the impact hazard on Mars spacecraft is minimal (Farnham et al. 2014).

*Acknowledgement:* This research is supported by NASA through Grant HST-GO-13610 from the Space Telescope Science Institute. This research made use of Astropy, a community-developed core Python package for Astronomy (Astropy Collaboration, 2013).

**Figure Captions:**
Fig. 1. Top row: Co-added F606W images of C/Siding Spring. Bottom row: Corresponding $1/\varrho$ enhanced images reveal the jet-like features in the inner coma. Celestial north is up and east to the left.
Fig. 2. (a): Color slopes of C/Siding Spring's coma measured from annulus apertures. (b): Maps of color slope (top row) and the corresponding $1/\varrho$ enhanced images overlaid with the contours of the color slope maps (bottom row). The color maps are computed with 5×5 pixel bins, and smoothed with a $\sigma$=2.35 Gaussian kernel.
Fig. 3. Zero-velocity syndynes for $\beta$=1.0 (blue), 0.1 (green), 0.01 (red), 0.001 (cyan), and 0.0001 (pink), overlaid on the F606W images. North is up and east to the left.
Fig. 4. (a): Half great circles corresponding to the two features. The filled circles are the position of the Earth as seen from the comet. The half great circles north of the Earth position correspond to the northwestern feature, and those to the south, to the southeastern feature. The widths of the great circles result from the uncertainty of the PAs. (b): Great circles defined by the bisections of the two features.

**Table 1:** *HST* observations of C/Siding Spring and the measurement results.

| UTC | Heliocentric [AU] | Geocentric [AU] | Phase [°] | Filter | Flux[a,b] [$10^{-15}$ W/(m$^2$ µm)] | Mag[a,b] | Slope[c] | $A(\theta)f\varrho$[a,b] [cm] | $A(0)f\varrho$[a,b] [cm] | Color[d] [%/100 nm] | Jet PA[e] [°] |
|---|---|---|---|---|---|---|---|---|---|---|---|
| October 29, 2013 | 4.582 | 4.042 | 11.1 | F606W | 4.68 | 16.98 | 1.03 | 1660 | 2520 | 5% | 156, 302 |
|  |  |  |  | F438W | 4.56 | 17.91 | 1.05 | 1550 | 2350 |  |  |
| January 21, 2014 | 3.768 | 3.686 | 15.1 | F606W | 6.26 | 16.67 | 1.05 | 1250 | 2120 | 6% | 168, 301 |
|  |  |  |  | F438W | 5.95 | 17.62 | 1.08 | 1140 | 1940 |  |  |
| March 11, 2014 | 3.279 | 3.793 | 13.8 | F606W | 6.22 | 16.67 | 1.10 | 1050 | 1720 | 9% | 133, 306 |
|  |  |  |  | F438W | 5.80 | 17.65 | 1.12 | 892 | 1460 |  |  |

[a] Flux and $A(\theta)f\varrho$ are measured from a 5000-km radius aperture.
[b] The uncertainty is about 5%.
[c] Slope is defined as $-\log(B)/\log(\varrho)$, where $B$ is the azimuthally averaged surface brightness of the dust coma, measured 500-5000 km from the nucleus.
[d] Color slope is measured at ~5000 km distance from the nucleus.
[e] Two numbers for the southeastern and the northwestern features, respectively. The uncertainty is 1°-2°, but 5° for southeastern feature at the first epoch due to its obvious curvature.

10Tricarico, P., Samarasinha, N.H., Sykes, M.V., et al. 2014, ApJL, 787, L35
Tsiganis, K., Gomes, R., Morbidelli, A., et al. 2005, Nature, 435, 459
Walsh, K.J., Morbidelli, A., Raymond, S.N., et al. 2011, Nature, 475, 206
Weiler, M., Rauer, H., Knollenberg, J., et al. 2003, A&A, 403, 313
Ye, Q.-Z., & Hui, Man-To 2014, ApJ, 787, 115
Yelle, R.V., Mahieux, A., Morrison, S., et al. 2014, Icar, 237, 202
10

**Figures:**

Fig. 1

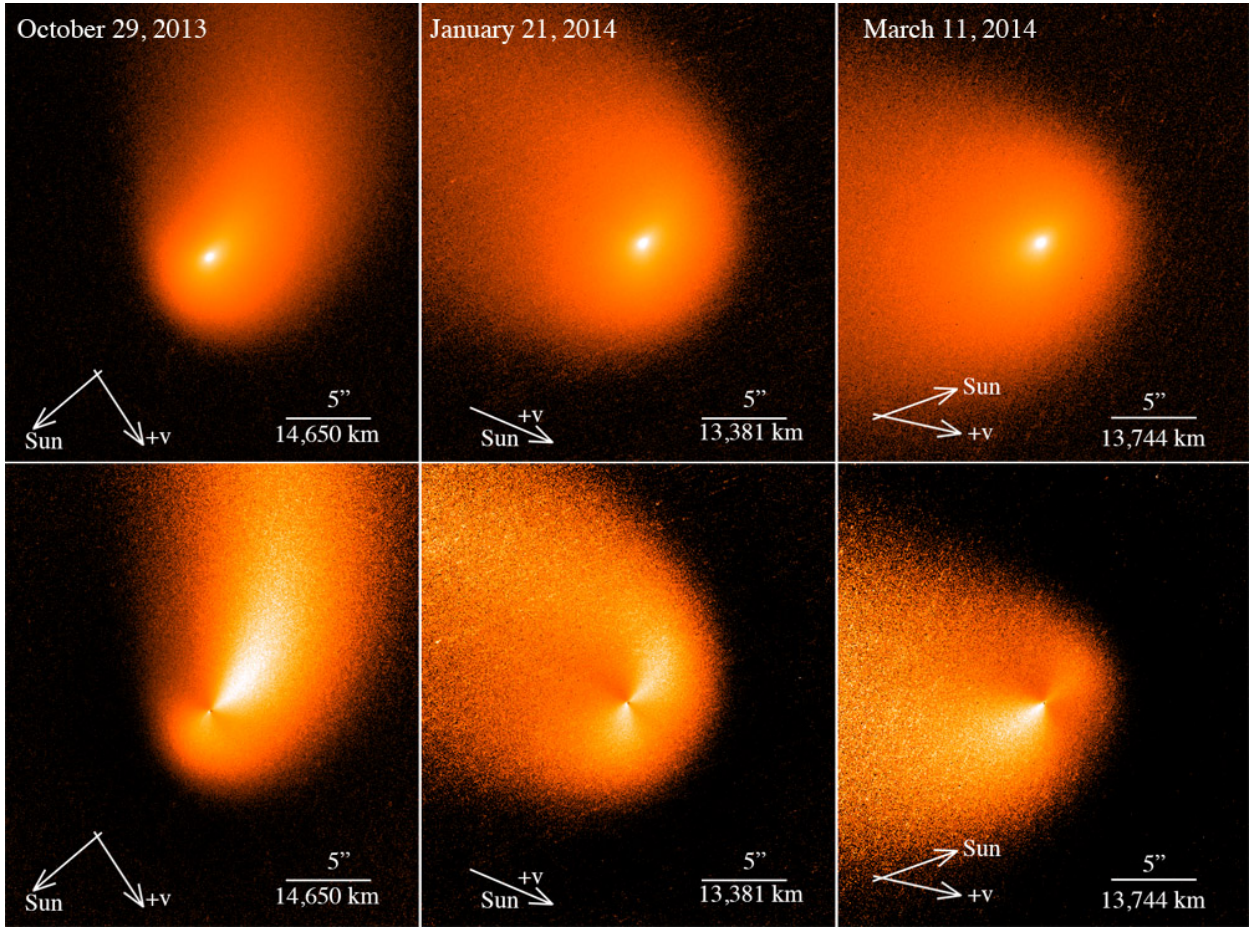

Fig. 2

(a)

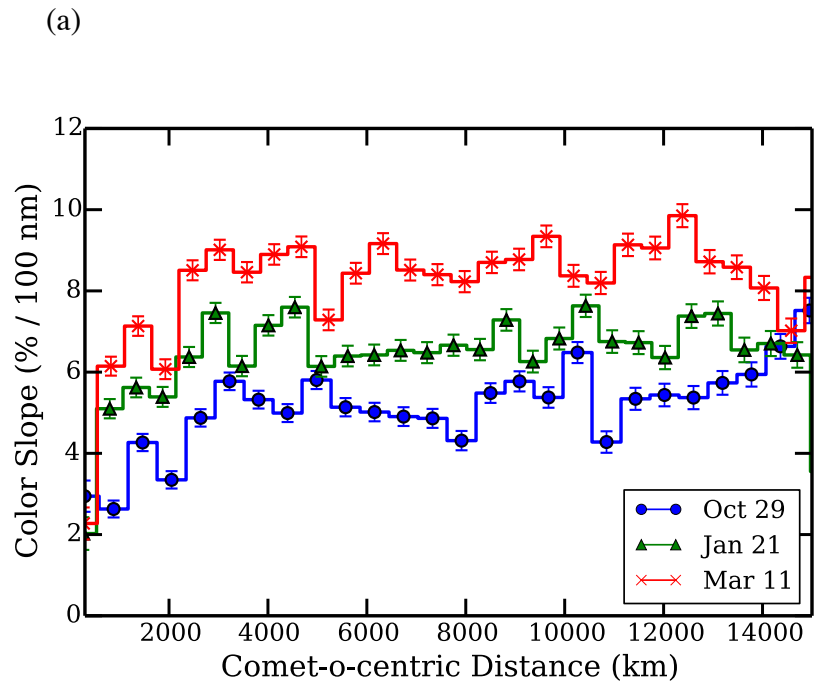

(b)

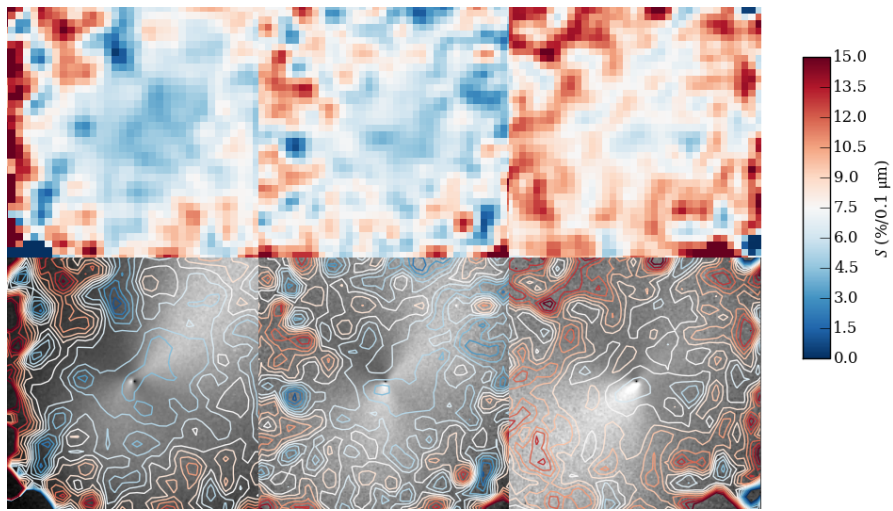

Fig. 3

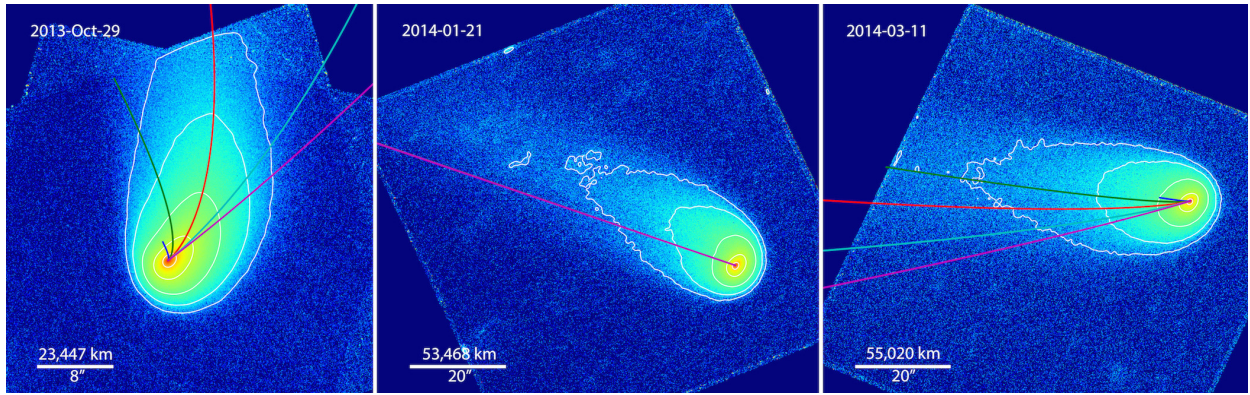

Fig. 4

(a)

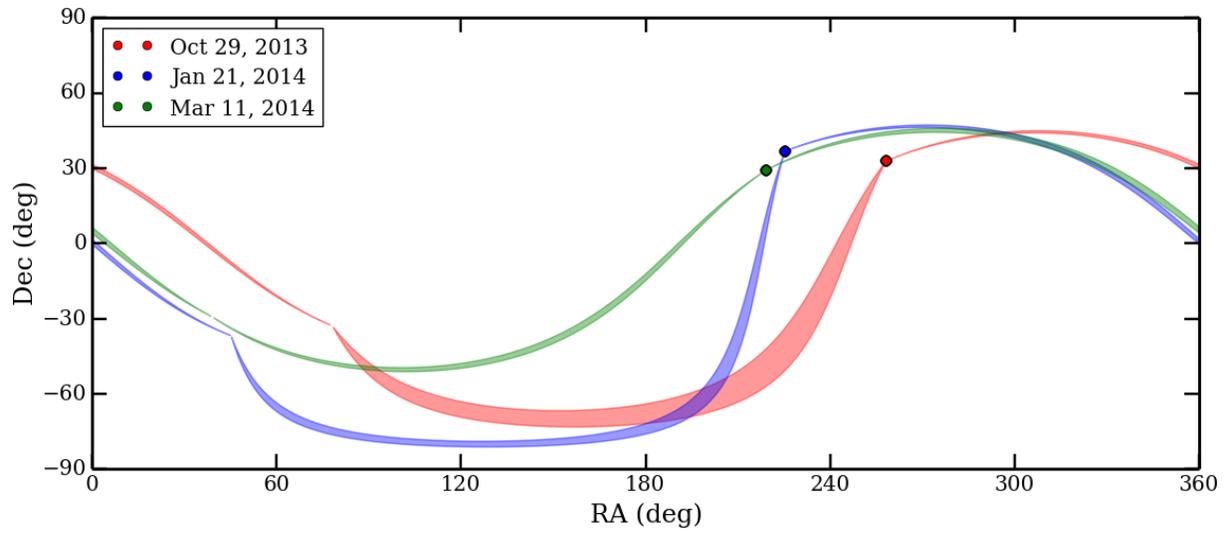

(b)

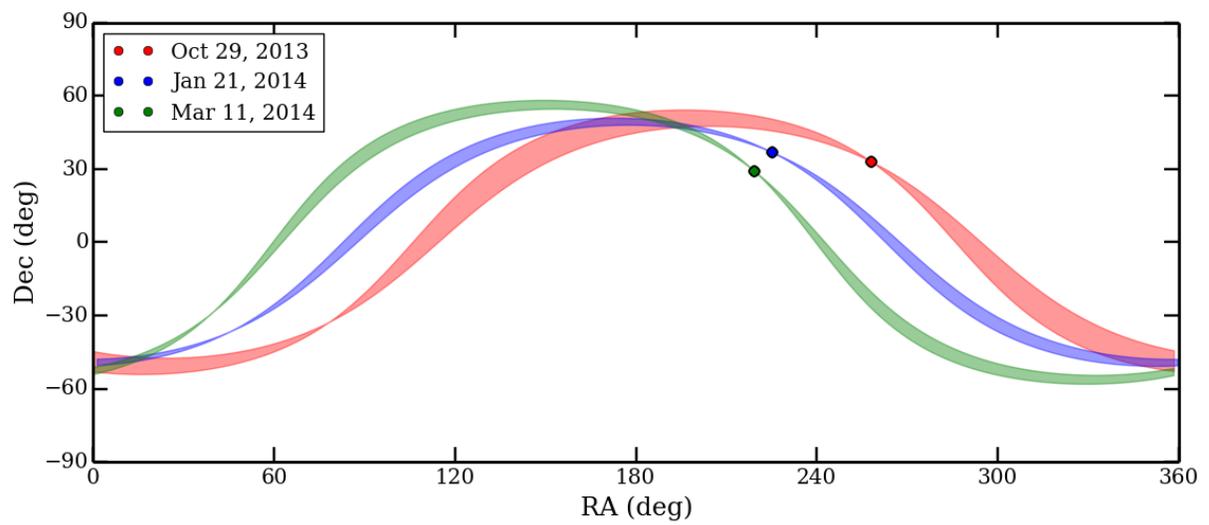